# THE IMPACT OF ARTIFICIAL INTELLIGENCE ON TRADITIONAL ART FORMS: A DISRUPTION OR ENHANCEMENT?

Viswa Chaitanya Marella

Department of Computer Science and Engineering, Vellore Institute of Technology
Vellore, Tamil Nadu, India
viswachaitanyamarella@gmail.com

Sai Teja Erukude

Department of Electronics and Communication Engineering, Bharat Institutions
Hyderabad, Telangana, India
erukude.saiteja@gmail.com

Suhasnadh Reddy Veluru

Department of Computer Science and Engineering, Vellore Institute of Technology
Vellore, Tamil Nadu, India
suhasnadhreddyveluru@gmail.com

**Abstract:**

**The introduction of Artificial Intelligence (AI) into the domains of traditional art (visual arts, performing arts, and crafts) has sparked a complicated discussion about whether this might be an agent of disruption or an enhancement of our traditional art forms. This paper looks at the duality of AI, exploring the ways that recent technologies like Generative Adversarial Networks and Diffusion Models, and text-to-image generators are changing the fields of painting, sculpture, calligraphy, dance, music, and the arts of craft. Using examples and data, we illustrate the ways that AI can democratize creative expression, improve productivity, and preserve cultural heritage, while also examining the negative aspects, including: the threats to authenticity within art, ethical concerns around data, and issues including socio-economic factors such as job losses. While we argue for the context-dependence of the impact of AI (the potential for creative homogenization and the devaluation of human agency in artmaking), we also illustrate the potential for hybrid practices featuring AI in cuisine, etc. We advocate for the development of ethical guidelines, collaborative approaches, and inclusive technology development. In sum, we are articulating a vision of AI in which it amplifies our innate creativity while resisting the displacement of the cultural, nuanced, and emotional aspects of traditional art. The future will be determined by human choices about how to govern AI so that it becomes a mechanism for artistic evolution and not a substitute for the artist's soul.**

**1. Introduction**

Artificial Intelligence (AI) is increasingly part of the process of making and curating art. From algorithms that make paintings and music, to robots that dance, various forms of AI technology are becoming a part of the visual arts, performing arts, and craft arts. This transition has raised questions about whether AI technology is a disruption to traditional forms of art or whether it is an augmentation tool that enhances human creativity. On one side are those who will argue that the artwork produced by AI lacks the authenticity, culture, and emotional depth that human-made art has [6]. While on the flip side are people who view AI as eliminating boundaries in terms of artistic expression and productivity [3, 8]. This paper will provide a full examination of AI's potentially dual roles in painting, sculpture, calligraphy, dance, music, and craft arts. We will look at evidence of both a disruption role and an augmentation role, case studies, and examine the technical and ethical implications, and conclude with possible outcomes to integrate AI with traditional artistic practice. To do all this, we will give you a well-rounded, academically sound exploration of whether AI is eroding very old traditions in art or ushering them into a new epoch.

## 2. Background and Technical Foundations

"Artificial intelligence" in the arts means many different things, but essentially refers to computational systems that can perform capabilities that mimic or augment actions that humans typically take through a creative process (such as learning and practicing to develop an artistic style, generating a new creation, or interpreting data artistically). In recent years, advances in machine learning, especially deep learning, have provided increased opportunities for AI to take a dataset of the visual arts, music, or dance, and to create new works in response to input, often with dramatically increased realism and authenticity [4]. Currently, there are a range of technologies being developed using generative adversarial networks (GANs), which match two neural networks against each other to create new images or music that is realistic but not identical, and diffusion models that take random noise and use repeated iterations to transform it to a coherent image [4]. Due to advances in transformer-based large language models, as well as other techniques, text-to-image generators (e.g., DALL-E, Midjourney) and text-to-music systems were developed that create artworks based upon natural language descriptions [3, 7]. In this context, so-called traditional art forms include both visual arts (painting, drawing, sculpture, calligraphy, etc.), performing arts (music, dance, theatre, etc.), and craft arts (handicrafts, textiles, etc.). The role of AI will vary considerably across these art domains. In visual arts, AI can learn the features of paintings or calligraphic scripts and produce new images based on those learned styles. For music, there are AI algorithms for composing musical scores that can potentially produce melodies or even entire scores, often indistinguishable (to untrained ears) from human compositions. In dance and performance, motion-capture data and choreography algorithms allow AI to suggest and, in some cases, express physically based movements. AI's technology is also applied in art conservation, such as computer vision and machine learning, to support restoring a damaged piece of art, or archiving to sustain cultural heritage [9]. Overall, these are important technical foundations to explore AI as both a disruptive and enhancing influence. It is crucial to reiterate that while AI can exhibit levels of creative thought and creativity through a recombination of learned expressions derived from patterns or collections of patterns, it cannot, and it's important to clearly distinguish the differences between experiencing art and experiencing intention from an artistic or human perspective [4]. Much of the debate about limits in artistic realms revolves around this important distinction.

## 3. The Dual Impact of AI on Traditional Forms of Art

AI has a Janus-like influence on the arts, both disruptive and productive. On the one hand, AI is altering the status quo of centuries-old artistic practices and livelihoods, permitting concerns of authenticity, originality, and equity. On the other hand, AI is providing a set of powerful tools and technologies that open up a new range of creative expressions, efficiencies, and democratization of the art creation process. We outline the disruption vs. productive dual impact below.

### 3.1 *Disruption*

The disruptive impact of AI on traditional forms of art is evident in several ways:

*3.1.1 Diminished human authenticity*

Artists and critics often argue that an AI-generated work lacks the genuine human experience, soul, or spark that marks traditional art [6]. In this case, the AI-generated output itself may be visually appealing as the AI image generators draw on datasets of human-made art and generate similar content that, while seemingly pleasing, lacks the lived cultural experience, reflection, and emotional struggle that goes into an artist's work [4]. Art is not just a product of aesthetics, but also a reflection of culture; indeed, some scholars claim that while AI-generated pieces may be novel, they will ultimately be imitative in originality, and we should hold art to be works that are made intentionally and knowledgeably to an audience [4]. This perspective asserts that AI-generated pieces, even if impressive, constitute aesthetic productions, but are not "art" in the richest sense - this distinction risks diminishing the idea of art as a communicative act made by humans.

*3.1.2 Plagiarism/Ethical concerns*

The generative AI system has been trained on millions of previous artworks and recordings, and frequently without the consent of the original artist. It has widely been claimed that AI art is fundamentally collage, or worse yet, so severely 'plagiaristic' - 'steal' the style and motifs of artists historically without crediting the original artist [4]. A group of artists recently contested that their copyrighted illustrations were trained on a popular AI image model's training set without consent in 2023 [6]. These examples raise deep ethical and legal questions. Many artists view unauthorized AI training as a violation of their intellectual property and moral rights. Lawsuits have already alleged that generative art models are infringing copyright by incorporating and replicating protected elements of human artwork [1]. Additionally, as AI can now produce art in the style of a living artist, some worry of reputational harms or dilution of the artist's brand; viewers pan during art consumption and use, may not distinguish, nor value the human artist with the same regard – if an AI uses the same mark [4].

*3.1.3 Decline in perceived value and skill*

The use of AI involving art likely can help create or strengthen the public perception that art is becoming an automated commodity rather than the fine art of skilled craft. Empirically documented, audiences develop a perception that if they later discover an artwork was co-created, along with AI, they may devalue it (and the human author) [10]. A recent study on visual art discovered that the incorporation of generative AI into the production process structured the valuation of the artwork and the artist followed [10]. The devaluing effect is the opposite of a viable collectible market among fans that disrupts the art market and artist livelihood with a collector's unwillingness to pay excessively high prices for a product that requires less input from "human" guidance. Similarly, traditional skills (life drawing, musical virtuosity, etc.) may be rendered irrelevant if the AI systems produce outcomes of sufficient quality with very little user effort involved. A younger artist may not feel motivated to properly acquire difficult skills, as legacy techniques may seem irrelevant, and the experience of working with an algorithm may seem like an easy alternative that produces similar visual outputs.

*3.1.4 Job displacement and industry disruption*

The most straightforward disruption that generative AI poses is to jobs in creative industries. Only recently, these generative systems have begun performing distinctly human tasks involving artistic production, including illustration, logo design, background score composition, photo editing, or even written content creation. For example, at animation and game studios, AI image generators can now produce first-pass concept art for everything from character design to props in a matter of seconds. This reduces the need to hire large teams of commercial illustrators for projects. A 2024 industry report outlined that about 200,000 jobs in the entertainment sector are likely to be materially disrupted by generative AI technologies over the next 3 years [2]. In addition to job replacement, these tasks involve compression of jobs (one person can create illustrations with AI tooling that would take several illustrators to complete), elimination of some jobs, and creation of new jobs that have not previously existed. Artists using traditional forms of media, such as painters, sculptors, and artisans, may be able to earn a living while exploring art, however, they may find less work available if galleries and companies replace living artists with relatively cheaper AI mediums based on faster output. The systemic impact could be uncompromisingly dire, from artist economic precarity to a consilience of creative output as the spectrum of human difference is replaced by algorithmically generated output. Stagnation of Creative Evolution: While AI offers endless recompositions of existing arts, critics worry it could lead to stagnation of creativity long term. AI is inherently retrospective learning – it learns through "reverse engineering" based on what came before. If AI is relied upon too heavily, it might promote derivative works that mirror past styles (because AI is optimizing based on the example it is trained on), not necessarily in a bold new direction (forward). Some artists fear a feedback loop, whereby new art continues to be based on AI's pastiches of old art, leading to a lower degree of originality over time. In a worst-case scenario, if future artists are significantly reliant on AI output, the act of taking creative risks and exploring new techniques could decline. This "regression toward the mean" effect would seem to be disruptive for the ability of the cultural evolution of art to take place. In preliminary evidence from an online one-off art platform, we see that while artists who use AI to create their work do seem to increase novelty of content on a one-off basis in the short term, this is not sustained in the long term (compared to artists not using AI) [3] - next to possibly most of AI usage leading to convergence on popular styles or motifs. These preliminary findings suggest that indiscriminate leverage of AI might lead to a homogenized creative process for many works, gravitating toward whatever patterns the algorithms favor, and this is problematic for traditional arts where human-driven innovation is paramount. There just does not seem to be a valid argument or formal ethical basis behind the potential stagnation of the very experimentation that pushes art toward growth and transformation.

To summarize, the disruptive effects of AI focus on undermining human aspects - ethically, economically, and creatively. The act of questioning authorship and originality is a short step away from removing the role of creative jobs (this is perhaps more geared towards the concept of developing work, too). Therefore, AI can deconstruct the way society commemorates and values "art" as we are aware of it currently. In any case, it is easy to call for caution about the adoption of AI in the arts as well as bolster the argument that AI presents more of a risk to traditional formats of art as we know it still today.

**3.2** *Enhancement*

While disruptions abound, AI also enhances and enriches the arts in many ways by enhancing what human creators do rather than replacing them. Some of the key enhancing impacts are:

*3.2.1 Enhancing Creative Productivity and Experimentation*

AI can act as a creative partner or an intelligent tool that enhances an artist´s capability. Studies have shown that artists who incorporated AI into their work can create more and experiment with more ideas more quickly [3]. For example, text-to-image generators allow visual artists to prototype ideas quickly because they can produce multiple candidate images based on simple text prompts, compared to developing simple sketches (days of work). In a 2024 empirical study, artists who utilized generative AI reported a 25% increase in creative productivity, on average, as well as more positive audience feedback on their art-sharing platforms [3]. The AI acts as a "co-creator" that handles the boring parts (rendering and testing out many variations) while human artists can start from the best ideas. Instead of limiting the artist, AI allows artists to access new styles or forms they might not have used on their own. The relationship between human and AI has also been called "generative synesthesia" [3], since human imagination guides high-level concepts and curation while the AI provides all the rapid visual (or musical) instantiations of ideas. In this way, AI is an enhancer of creativity; it expands the artist's toolkit that has powerful generative capabilities and can even be a catalyst for new creative strategies that synthesize algorithmic and human aesthetics. Democratization and Accessibility: Artificial intelligence allows for lower barriers of entry into the arts, making it possible for more people to participate in creative expression. For example, a person with no painting training can take an AI image generator and describe their intention in words, and the AI creates the image. Someone who does not play an instrument can hum a tune to an AI software that can complete the piece into a full composition. This democratization enables more voices and perspectives to intervene in the arts, and thus, potentially increases diversity in artistic outcomes. Similarly, AI is creating easier access to participation in the arts and arts education. Turning our thoughts to the performing arts, and thinking about dance for a moment, we might consider that AI tutor systems provide personalized and real-time feedback to dance students anywhere in the world. One AI-enhanced learning platform is DancingInside, which employs machine-learning algorithms with pose estimation to illustrate similarities in a student's movements and a professional instructor, which produces action-specific corrective feedback [5]. A user study with dance students showed that when training with the AI tutor for two weeks, their dance ability exhibited refinement that closely resembled the degree of enhancement they would have received if taking lessons with a dance instructor in person [5]. This is an example of how AI can enhance access to a high-quality training experience without the constraints of access-boundaries based on location and affordability. It "breaks the geographical barriers" that are often associated with the provision of arts instruction [5] and delivers expert experience to remote or underserved communities. AI can also assist disabled artists as well - generative art programs can help artists who cannot manually paint and/or are unable to play an instrument to still use AI mediation to articulate their creative ideas. Ultimately, AI promotes access to the art world for new participants and audiences. New artistic forms and hybrid practices: DI uses existing art forms and supports the development of new forms and hybrid approaches. Generative art, wherein an artist develops an algorithm so that the algorithm moves from a fixed location to a new location autonomously to create images (or even music), is a widely accepted genre. The aesthetic present in AI-produced art is one of glitches, surrealism, and hyper detail, which provides a new domain of creative possibilities. Artists can work with algorithms as the automated creator, meaning that they are responding to human intention with the machine output. In the space of music, for example, composers have begun to use AI suggestion tools to create melodies or harmonies that they otherwise could not create and to use that output as their "musical muse." Some choreographers are collaborating with AI to produce movement patterns; a project building on their pioneering work demonstrated this by training a generative AI model on decades of dance videos to produce choreography that was interpreted and refined by the human dancers [5]. The hybrid approaches that can develop in procedural art can help extend boundaries of each art form, merging human spontaneity with algorithmic variability to create new performances and artworks. AI is also enhancing the audience's experiences of interactive art installations- for example, there can be AI-driven visuals responding in the moment to a musician at play or to a dancer at work creating a new kind of immersive art experience.

Improvements in preservation and restoration: About crafts and heritage arts, AI is providing paths to the revival and sustainability of traditional forms. The process of restoring paintings, sculptures, and historical objects has benefited from AI-based image analysis that can accurately predict original colors or patterns in damaged areas [9]. AI vision systems can consider an old fresco and digitally remove centuries of grime or insert missing fragments to help restorers. This helps me refinish restorations more faithfully and can help age-old artworks restore some of their original brilliance. Archiving and analysis are another realm - AI can digitize thousands of manuscripts or craft designs and categorize them, especially in the search for patterns of styles or lost techniques that a scholar or artisan may find useful. For instance, machine learning models have analyzed hundreds of types of calligraphic scripts to identify a specific master's hand in unsigned works or create new calligraphy in historical styles for educational purposes. These developments bolster the conservation of our cultural heritage not just as a matter of enduring tradition but as a representation of the ongoing vitality and renewal of traditional art through a contemporary lens [9].

*3.2.2 Economic and Collaborative Possibilities*

In addition to various cultural implications, welcoming AI can improve the socio-economics of the arts in positive ways. To be sure, while AI will enact some measure of automation, it opens up demand for positions: artists that are skilled in using AI tools are in high demand right now in design, gaming, and media businesses. New roles in art are being created by technology, such as AI art curator, prompt engineer (a person drafting the prompts to create AI art), or machine learning designer that incorporates both technology and artistry. These roles offer avenues for traditional artists to diversify their practice, rather than replace them altogether. Additionally, AI can take on many of the laborious repetitive production aspects (coloring animation frames, generating base instrumentals, etc.) and allow human creators to spend more time making higher-level creative decisions. This could mean efficiency of cost in creative industries as icebergs of creativity are unlocked, which increases encouragement to fund more projects (as there are reduced labor costs) and allows small artists to produce at a scale that would previously require large teams if using software support. At a collaborative level, if an artist, for example, a painter, were able to utilize AI to repurpose her painting into a musical soundscape, she can connect with a 'digital composer' without the two physically being together and then allow AI to do what it was trained to do. With interdisciplinary mashups, AI-enabled collaborations can breathe new life into creation and offer opportunities for the cross-fertilization of ideas.

In summary, I see that the enhancing aspect of AI lies in its ability to be a powerful new tool in the contemporary artist's toolset, and when used smartly, can fast-track human creativity, accessibility to art, and respect for artistic traditions. It doesn't denote a dehumanization of the artist, but rather a way to expand the art/cultural process. There exists a sort of duality, where AI can be both brush and muse, while expanding the horizon of possible outcomes for creatives and audiences. The next chapter will look at specific instances that illustrate disruptive and enhancing trends side-by-side.

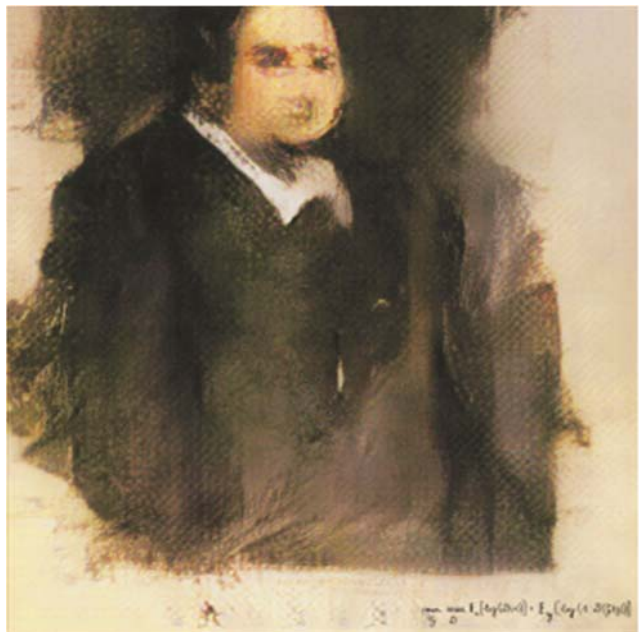

Figure 1: Edmond de Belamy (2018) by the art collective Obvious – one of the first AI-generated portraits to be auctioned in the traditional auction system. This GAN-produced portrait, developed by an algorithm trained using moments in history/time for artworks, sold for around 432,500 USD at Christie's. The blurry and unfinished end product is an intentional outcome of the AI method. The auctioning of Edmond de Belamy gave birth to worldwide questioning and debate about AI in regards to art and the boundary between human-made and machine-made.

## 4. Case Studies and Empirical Evidence

To demonstrate both sides of AI's impact, we will explore a few case studies from a range of domains in the arts. We will also include empirical evidence and actual examples.

**4.1** *AI-Assisted Illustration and fine art*

One valuable case study comes from an online art community where some artists began creating digital paintings with text-to-image AI (e.g., Stable Diffusion). A recently published, peer-reviewed study examined thousands of artworks uploaded to this platform and found significant increases in productivity and audience engagement for artists who used AI (3). These artists, on average, not only produced 1.25x more works already, but they received more "likes". Their audiences seemed to appreciate that their creative output had been augmented by AI (3). There was also a nuanced impact on creativity found in this study; although the average novelty of AI-assisted art declined over time, as artists began to rely heavily on familiar forms created using AI (3). Likewise, the maximum novelty (most original works) of the AI-art users slightly increased, which seemed to indicate that some artists used AI to push their imagination beyond their experience (3). This case demonstrates both enhancement (higher output; positive audience reception) and a certain level of disruption (convergence in style) overall in the practice of visual art. Another famous and documented case is the sale of an AI portrait, mentioned earlier, which took place in 2018 (Figure 1). The sale of Edmond de Belamy - a GAN portrait - for hundreds of thousands of dollars was empirical proof-of-concept that AI art can see success with recognition in elite art markets. On the other hand, it resulted in backlash from traditional painters and critics who wanted to know if the work was truly by the algorithm versus the human team responsible for the curation and presentation. This demonstrates the underlying tension: AI can produce works sufficiently impressive to rival human art (enhancement of art) but in doing so, opposes the norms of what is not art and authorship (disruption of paradigms).

**4.2 *Performing arts – dance and music with AI***

In dance, AI has been used in both training and choreography. One relevant example is the DancingInside platform (developed by researchers in 2023), which we briefly discussed earlier. In this case, it allowed a group of ballet students from geographically remote areas to train over weeks with AI feedback. The results were published in a paper: dancers using AI feedback improved their technique at the same rate as peers who were taking in-class classes [5]. Additionally, teachers reported that the AI could detect minor issues related to posture with pose analysis that a teacher might not notice in real time [5]. The utility of this case study comes from the demonstrated enhancing capability of AI by democratizing access to high-quality performing arts education. Another example is choreographer Wayne McGregor, who worked with Google's AI lab to make choreography using an algorithm to learn patterns of McGregor's style with motion-capture data [5]. The algorithm produced choreography with new sequences of movement, which McGregor then interpolated into a live performance with his dance company [5]. Dancers were able to (effectively) "co-create" moments with the algorithm's feedback, using the algorithm's suggestions as stimuli for human interpretations. The reviews commented that the results were within McGregor's style, but the way the transitions and formations were structured felt surprisingly new to them as an enhancement that the human–AI relationship provided. In music, AI-composed songs have begun to breach the threshold of mainstream production. For example, although the pop album Hello World (2018) was an early AI-human collaboration, the capacity of AI to generate music has developed in sophistication in more recent years. A user study in 2025 examined listener reactions to AI-generated music and human-created music pieces [10]. Interestingly, listeners frequently did not recognize the AI music in blind tests, and many claimed to like the AI pieces for their melody and mood. Still, when told that a piece was by an AI, some listeners attributed it to a "lack of emotional depth" – this may be attributable to bias, or perhaps the AI did yield a subtle quality gap [7]. In this performing arts case study, two points are demonstrated: AI can create and generate performances (including full performances where an AI generates the visual and accompaniment live), defining an expansion of scale and interactivity of shows; and AI creates a conversation about the authenticity of performance – should "best composition" go to a songwriter, or to the group or producer that pointed an AI to produce a hit song. Performing arts are experimenting with AI as a creative aid, while managing, negotiating, and debating disruptions to attribution and the nature of artistic performance. Traditional Crafts and Cultural Heritage: An interesting case at the crossroads of craft and AI is a project involving Guatemalan artisans reported on in 2023 (Camarillo et al. 2023). In this project, master woodworkers and metalworkers were trained to use generative AI tools (specifically image generators like DALL·E) to brainstorm new design motifs for their traditional crafts. The artisans would provide a description related to their cultural heritage (e.g., "Mayan-inspired door knocker pattern"), and the AI would provide design recommendations. Those AI designs were then physically created by the artisans in bronze door knockers (Camarillo et al. 2023). The findings, included in the publication "Modern Tools, Ancient Skills," showed an impressive improvement: the AI sped up the design process (a process that used to take up to another week sketching and prototyping was reduced to about three days), and excited artisans to follow new creative paths they otherwise had not imagined (Camarillo et al. 2023). One artisan remarked that the AI "rapped together motifs from different villages that I would not have thought possible to combine" - this demonstrated an expanded creative capacity. The project also revealed limitations: some artisans expressed concern about relying too much on the AI and whether using the AI would slowly chip away at their traditional aesthetics or skills, which had been passed down from previous generations (Camarillo et al. 2023). There were also questions surrounding cultural authenticity - if an AI system suggested a design, could we still consider it as "authentic" to the culture, or does it dilute the lineage of the craft? The findings of the case study aptly distill the duality of AI's impact: the efficiency gained and the innovative cross-pollination of ideas (benefiting both the craft business and creative portfolio) against the concern for integrity and identity of traditional forms of art (potential disruption of heritage). An additional case in preservation is when AI was used to restore ancient murals in Asia - heritage conservators used neural networks to take educated guesses to extrapolate missing parts of centuries-old murals, based on conventions of learned artistic practice [9]. Such AI-assisted restorations were able to reconstruct plausible images of faces and scenes where paint had worn away, thereby enhancing the completeness of the visual story for museum visitors. This shows the value of AI in cultural heritage- it can act much like a time machine, reconstructing art that would have otherwise remained obscured, thereby maintaining the continuity of traditional art practice. The disruptive aspect here remains small, aside from the discussion on how far the AI should go with its creative license for guessing historical art (it's up to museums to decide what is original vs AI). Overall, the crafts and heritage case study examples demonstrate AI, when used carefully, as a pathway between the past, present, and future: AI can breathe new life into traditional arts and heritage, while conserving them for the future, all in the context of authenticity and respect for original art forms.

These case studies demonstrate that the impacts of AI on the arts are not some abstract future speculation; it is happening in real studios, real stages, and real workshops all around the world. The evidence provided presents a complex picture: AI acts as a possible driver for artistic innovation and cross-cultural exchange while igniting friction with existing artistic norms and economies. In the following section, we continue building off these examples to discuss the common challenges and limitations that arise, along with possible ways to address them in advance, to maximize the benefits of AI and limit the harm.

## 5. Challenges and Limitations

While the prospects for utilizing AI in traditional art production are exciting, there are substantial challenges associated with the integration of AI into the traditional art sphere, from the technical limitations of today's AI models to major ethical questions and disruptive socio-economic impacts. We categorize challenges into technical, ethical, and socio-economic challenges, with the understanding that they are interrelated.

### 5.1 *Technical limitations*

The recent advancements in AI systems are significant, but many of these models and systems suffer from technical limitations that shape their role in the arts. A glaring limitation is the complete absence of true creativity or comprehension. AI models lack consciousness, subjective states, emotional, or cultural history; they utilize pre-trained data to create patterns. Sometimes, AI-generated art can be impressive and initially exciting, but on closer inspection may feel superficial, hollow, or formulaic [7]. For example, AI-created music could recognize harmonies and follow rhythms, yet emotionally leave the listener empty, especially when a genre requires intense storytelling or personal reflection. A robust 2023 review of music AI generation noted, "AI outputs were often predictable and repetitive, which nixed the emotional depth and originality that human composers can bring to music." [7]. In summary, the very thing that limits AI in producing works for the arts is the inability to produce works in an autobiographical context and without intentionality; AI creates variations of what it has seen, but cannot intend meaning as an artist can. Just like in visual art, AI-generated images may exhibit an eerie sense of coherence, but they may also exhibit errors of subtlety or nonsense (e.g., having an extra finger in a painting) because the AI does not have a true grasp on what it is reproducing, as it lacks a concept model that a human brain can utilize. Another technical aspect is the issue of control or bias. When artists work with AI, they often find themselves not in control of the fine precision in the AI's output performance. You can prompt an image generator with a style or subject, but it will not precisely turn out how you envisaged, nor will it be entirely predictable and fine-grained to steer the AI towards a specific vision. The fact that deep learning models are "black boxes" means that when the artist interacts with the AI, it is generally a process of trial and error, which can be cumbersome and frustrating. Particularly, there are times when the artist must adjust their vision based on what the AI is capable of producing, which can curtail true artistic freedom. Furthermore, AI models exhibit bias that is rooted in their training data, which means if datasets have gaps or unequal representations (e.g., more western art than non-western art), those biases will be exhibited in the AI outputs and create suboptimal or a lack of diversity outputs. For instance, an artist may wish to create a landscape using a landscape generator trained mostly on European paintings, and is expecting an East Asian ink wash style landscape, however, the generator may not produce an authentic landscape retaining its East Asian provenance. This type of technical bias may inadvertently de-prioritize certain artistic traditions and could be viewed as a limitation or potential stereotype when looking to use AI about the Arts internationally, unless the artist re-trains the models or is screening their datasets. From a performance-oriented view, latency and accuracy are additional, simple but practical technical issues. In vivo situations (for example, an interactive AI dance in performance), the AI needs to take in input and provide output, in real-time. If the AI requires high computation, it may exhibit some latency, which would disrupt the performance flow. Similarly, if the AI misperceives a particular performative action (for example, when it takes a dance gesture as something entirely different), it may provide incorrect visual or aural stimuli, disrupting the performance. Overall, there remains a serious technical challenge in reliability and the real-time engagement that something like this entails. A third issue is the scope and cost of these models. The leading AI models (specifically the newest generative transformers or diffusion models) utilize vast computational power and training and operational energy requirements that are significant to run. This may be simply unavailable to independent artists or small cultural organizations, who may consider the use of these models to be prohibitively expensive. Further, the quantity of resources required raises questions of sustainability, precisely in our climate emergency, as training many large-scale AI models has a carbon footprint. All of these issues may prevent equitable access to the use of AI for art-making, concentrating the technology on large tech companies or highly-resourced studios, rather than democratically amongst creators. In conclusion, although many of the technical restrictions encountered loom large in the arts practice of regularly using AI, these problems are being addressed in research. Ultimately, AI will still be unable to exhibit the real creativity and fine control (and sometimes speed as well) that a human artist can in the art-making process, and it will not replace the honed skill of a human practitioner. AI can be best thought of as an assistive tool that does not know, or envision, or know what a "good" outcome is, and that ultimately still relies on a human to imagine and to adjust the weirdness that unavoidably will accompany it in the use of AI.

**5.2** *Ethical issues*

There are various ethical issues regarding the AI-arts interface in practical use, and many legal questions have not yet been settled. One of the main ethical issues is the issue of authorship, copyright, and ownership of work produced by AI. Copyright law, as it applies to humans as creators, is starting to be challenged by AI in many jurisdictions. In 2023, in a landmark decision in the United States, a federal court reaffirmed that a work entirely generated by AI is not copyrightable because it lacks a human author [1]. Similarly, the U.S. Copyright Office has promised to ensure that only protectable works, with human creative input, are copyright eligible [1]. There is, therefore, a grey area for art and contracts involving AI-generated elements. For example, if a human artist uses AI to generate elements of an artwork, which (if any) of those elements are copyrightable? Who owns them? The risk that any generation of art by AI cannot be copyrighted means it can be copied, with the loss of economic rights to the artist (or algorithm designer). Other people may suggest that legal protection could create an incentive to proactively manage AI with responsibility towards artists. On the flip side, delegating authorship to AI or its operators could materially disadvantage human artists, where the probability of holding thousands of images owned by an AI company would contribute to the marketable value with copyright. The consensus in ethics is still in formation, however, the law tends to choose a position that requires human creativity to generate legal authorship [1], and AI can be seen as a tool in that model. However, without a clear ownership model for AI creation, we run the risk of remaining "in a contract," without AI artist creators wishing to use AI for fear of not being able to take copyright protection, or exploitative practices, whereby corporations profit from commercializing AI art without compensating any contributing creators. Likewise, the issue of consent and fair use is another ethical consideration. Beyond the moral outrage, training on an artist's artistic data with no consent has led to lawsuits. Artists whose work is harvested from the web are concerned that, with the model built from their work, they are being removed from the market, entitled "learning from me to compete against me." Should artists have to consent or be compensated with a licensing transfer to have their works incorporated into any training set? Many in the arts are concerned that transparency should improve if there is an opt-in model for training data [4], and some AIs are beginning to allow artists to, voluntarily, remove from training corpora, but it's not normalized in the industry yet. Another serious concern is plagiarism: if the AI can generate artwork very akin to a living, working artist's style, or even uniquely specific painting (as demonstrated, in circumstances where the prompt created a specific copyrighted image), who is liable? The user, the AI? The dataset? Behaviours of ethical behaviour in the use of AI would limit this from happening, using alternative models could limit improvement by overfitting on one work. Where do moral rights fit in? Even if the use is legal, is it right to generate derivative works in the exact style of the artist, when the artist is stating clearly they object to these works? Many cultural actors would say that insistence violates the ideals of respect for the creative agency of the artist. Attribution is a related ethical question. When artwork is created with the (accumulated) works of artists incorporated into the process, what should attribution involve? The individual who created the concept and guided the AI deserves credit, to be sure, but should the algorithm (or the algorithm and the creators) receive credit? For example, if a photographer starts with a favorable black-and-white image and uses an AI tool to colorize the photo, the final product is two pieces: the original photograph and the AI's learned expertise from millions of colorized photographs. In terms of ethics, there could be a justifiable pro-transparency argument; artists should declare AI assistance for the sake of honest disclosure, so the audience is not misled. There has been a movement in art competitions and exhibitions to require artists to declare AI-assisted work. Ultimately, the disclosure informs the audience or judges on what criteria to consider (the audience may consider different expectations for a piece physically created vs. one involving code). The opposite side of the argument is that AI could very well become a very common tool (e.g., Photoshop) over time, and not every association comes with a declaration. This conversation is still evolving, but over the short term, it seems that artists and organizations will increasingly be pressured to offer transparency about AI, given the novelty and sensitivity that pervades the public discourse around AI in art. Talking about ethics raises more issues around bias and cultural considerations that must be acknowledged. AI systems are inherently flawed and soak up the socially established biases as they are trained to produce outputs. The possible outcomes can be easily based on the nuance of the underlying bias. For instance, a model trained primarily on white European Renaissance artwork could be based on very obvious triggers from other cultures' design, or even worse, become biased when creating art for a specific gender or religious theme based on stereotypes implied in the training data. A problematic outcome based on bias that would be viewed as art continues to perpetuate the cyclical stereotype of harm. From an ethical perspective, project creators should be deeply aware of these considerations and should avoid these pitfalls as part of their practice in producing AI art. Moreover, considering that AI tech is globally collected, one would imagine that one group could utilize AI to copy a style from another culture into the new positioned expectation without an understanding of the importance of the work or subject matter (for example, an AI generated a sacred indigenous art pattern as decoration). The ethical considerations also go into issues of cultural appropriation, which some indigenous art communities worry that artificial intelligence copied their traditional designs, and if misused, are not held accountable. Ethical considerations should include working with communities, establishing boundaries for datasets and designs for examples that would need a consultation approach, and making sure to build AI with considerations of respecting cultural perspectives. Overall, the ethical considerations are just as hard a problem as the technical. Being fairly compensated for work and obtaining consent, establishing or reclaiming authorship or intellectual property for AI art, and recognizing issues or unintended harm through biased AI outputs are real issues that must be addressed. The art community and policymakers are just beginning to grapple with the questions. A good response would require revisions to the existing legalities, possibly rethinking how the industry would know its responsibilities, or even a healthy strategy for discussing these issues among technologists, artists, and ethicists.

### 5.3 *Socio-Economic impact*

AI's rise in the arts has broad socio-economic implications that include artists, institutions, and the wider society as a whole - it does not happen in a vacuum. The most immediately pertinent is the impact on artists' livelihoods. As AI tools become more powerful, there is apprehension that artistic professions will diminish in number. For example, concept artists and illustrators who operate in fields like publishing, advertising, and game design are already having their roles competed with by generative models that can make visuals for a fraction of the cost and time. A company that once employed several junior artists to format and draft storyboards will no longer require even one storyboarding artist, only an art director with an AI generator. This aggregation means fewer entry-level opportunities for emerging artists; the arts career landscape may become even more unstable. Likewise, background music composers and session musicians will see gigs dwindle if studios prioritize AI-composed background scores instead. The Animation Guild's 2024 report imagined a shift in fun indentation where the effects of entertainment industries create significant labor unrest to the status quo: merging job roles into fewer roles, new technology-based roles, while many more roles simply disappear. It warned that creative workers will be "facing an unstable era of disruption… characterized by some roles being merged, new roles being assigned to replace existing roles, and many roles being eliminated." In terms of economics, this indicates an opportunity for debt servicing and upskilling: artists will need to be willing to learn to use AI, or be replaced by it; artists will become as technologists as they will creators. Flipping the job loss on its head is the risk of new markets and opportunities. Since AI can lower production costs, it's conceivable that there could be more content created in general, potentially leading to more aggregate demand for artistic work. For instance, if a filmmaker can use AI to do visual effects, then more films get produced, thus employing more actors and writers, etc. And artists who become adept at using AI could see consulting work or opportunities with other types of work, like curating AI outputs or specializing in combining crafting style with an AI art style for a higher price due to its uniqueness. Additionally, AI can facilitate the entrepreneurial potential of artists, too; so products could be designed entirely by a solo craftsperson with the help of AI (design, branding, marketing visuals) that might usually require hiring several people specifically for those tasks. This productivity boost could lead to increased income for those able to utilize this tech. There's also a socio-economic divide to this. Advanced AI tools and infrastructure are expensive - well-resourced organizations (big studios, huge tech companies, rich country museums) will be able to adopt new AI tools faster than small galleries or artists in developing countries. There could thus be a widening inequality in the art world, where those who have access to AI will grow faster in productivity and reach, while the others will be further behind. The democratizing effect could a year or two down the road, be that the tech will be cheaper (as it does eventually trickle down) and thus will help mitigate disparity, but right now, this is worrying. It highlights the need for AI tools to be democratized and training to be democratized, because if not, benefits will only go to the privileged few groups. From the perspective of cultural institutions (e.g, museums, galleries, performing companies), AI is raising questions about public perception and the value of the arts. If there is an overproduction and saturation of artifacts in the market, will the audience pay for performances of human orchestras or pay for original paintings? Or, will there be a counter-movement valuing "hand-made" works, similar to how the art of hand-made crafts commands their value in a mass manufacturing world? We are observing some pushback already: for example, there are art communities and art competitions that have banned AI-generated submissions to preserve human-only spaces for creativity. The socio-economic consequences in the long run could lead to art splintering markets or audiences - one for AI-supported or mass-produced creative content (often free or low-cost, primarily for background entertainment, etc.) and one for human, artisan-crafted artwork that is valued accordingly (more expensive and exclusivity). Artists could pivot to speak to narratives and imperfections on where the artist was present, and those different selling points that AI can not touch. Finally, there is the social problem of how the role of the arts will change. In particular, the arts have historically been a prominent means for human expression and commentary, reflection, and to elicit feelings. If AI-generated substances become ubiquitous, will all art be commodified? That is, art will be consumed more like an on-demand utility (e.g., generate music for every mood, or generate AI visuals for every playlist) rather than a deep interaction or relationship? And, this has implications for society's consideration of how it will fund the arts - that is, if as a taxpayer sees that AI created sufficient art for cheap and quick consumption, why would a taxpayer support arts education, or provide grants for artists that the taxpayer does not see the intrinsic human value around? It will be a challenge for the arts community and advocates to describe the continuance of human creativity in the world of AI. Ultimately, the socio-economics of AI in the arts appears to have a double-edged sword: increased efficiency and expansion on one side, and displacement and inequality on the other. To deal positively with this will take some decisive action: all-encompassing retraining programs for artists to re-learn AI, create new economic models (possibly artist royalties for AI companies who use their work to train), and public policies to establish and preserve the arts as a profession. This will be expanded upon in the next section, which will explore some potential future directions and recommendations for organizations and practitioners to ensure AI is used as a tool for enhancement rather than disruptive detriment to the art and cultural landscape.

## 6. Future Directions and Recommendations

In considering future direction, the goal is to take advantage of the benefits AI has to offer to the arts while mitigating any potential pitfalls. This will require a multi-faceted approach with different players in the AI ecosystem coming together: technologists, artists, educators, and the government. The following recommendations and future directions are important to assist in establishing complementary relationships between AI and traditional art forms:

### 6.1 *Adopt human-AI collaboration models*

Rather than positioning the narrative as AI versus artists, the future must focus on AI with artists. AI literacy should be included in educational programs (not just computer science) in art and design schools to teach emerging artists to use these tools creatively and critically. This will prepare artists to take on the role of "AI director," where they guide algorithms to serve their professional vision, intentions, or agenda. Collaborative creation should be actively promoted—altogether—competitions or festivals specifically for human-AI co-created art could be used to establish a new hybrid art form. When we consider collaboration, we recognize/check that AI acts as an assistant or creative partner, not a replacement. This collaborative mindset can slow the polarization of "AI vs. art" and ensure that artists remain open to experimenting with AI without fear of the unknown.

### 6.2 *Create ethical principles and regulatory guidelines*

There are many questions to be answered to provide clearer guidelines on consent, attribution, and fair use of artistic creative data. AI companies and communities would be well-suited to create something along the lines of "Artists' Bill of Rights" regarding AI training datasets, with the intent of securing rights for authorship. The goal is that artists, whose works substantively inform an AI model, are either asked to provide consent, receive credit in the model documentation, licensing framework, or receive compensation (a licensing or profit-sharing agreement with AI companies if commercial purposes are intended). From a legal standpoint, copyright law may need revision to account for AI about authorship in given works. A suggested mechanism could be that we allow copyright for a finished AI-assisted product, as long as the human was able to demonstrate a significant amount of creative contribution (this could allow for the protection of artists who incorporate AI into their studio practices). In addition, we could create new categories such as "AI-generated" for transparency on AI assistance, without implying full characterization of authorship. Policymakers could convene with technologists and artist restructuring bodies to collectively establish balanced policies that protect the interests of human creators while providing sufficient protections for creativity. Likewise, there certainly must be international dialogue between jurisdictions, as both AI and art cannot be contained within borders, and an international standard or at least interoperability will limit the ability for someone to be able to use regulatory arbitrage.

### 6.3 *Collaborate on improving AI tools with artists*

AI tools for creating art in the future should be developed collaboratively with artists to develop tools that are created with a specific intention for an artist and cultural perspective. For instance, providing features for style attribution (where an AI could be instructed not to closely copy a particular artist's style) could preempt a lot of concerns around plagiarism. Research into explainable AI for creativity may lead to creators being able to understand the reasoning behind how the AI arrived at a conclusion that led to the result, making the product neither an entirely ambiguous entity. In addition, the incorporation of broader and inclusive training data, with opt-in contributions from artists from around the world, can make the AI model more "tasteful". However, it is of significant import that we work with artists when developing AI, potentially as part of studio residences, to help ensure that the future directions of creative AIs are in alignment with the ways that artists want creative AIs to think and work with, as opposed to just the way its technically possible.

### 6.4 *Encourage equal access to technology development*

We have to make sure all artists benefit from the advantages of AI in art, not just artists from privileged worlds with access to tech incubators or wealthy institutions. We could advance open-source artistic AI tools, affordable or subscription-based services for individual creators, and workshops to enable artists from communities that don't necessarily have access to the arts sciences. In one example, an initiative called for "inclusive technological development and [encouraging] dialogue between traditional and contemporary practice" [8] regarding Guatemalan artisans: in this case, technology-regardless of its method- but by actively work with the technology, the technologists needed to work with them to adapt the technologies needs without removing cultural identity. For developing countries, we may be able to have software developers partnered with craftsmen or dancers, or musicians to develop local AI technologies that enhance their artistic forms. The point is we will have no idea how to use AI technologies for art. To prevent a situation where AI arts become elitist, democratize the technology so that it is available to lend a hand in changing the arts landscape.

### 6.5 *Advocate the significance of human artistry*

The arts community - presuming that educators, critics, and artists themselves will need to create an environment in which we openly discuss what's great about human art; what story is behind the emotional states and context of human works that an AI cannot imitate. This should not be done in a dismissive way to AI arts. Possible future directions include that museums and galleries could create exhibits comparing AI and human-produced that are not competitive but instead provide a discussion so that audiences can appreciate both for what they are. Emphasizing process, such as illuminating how a human process translates into a painting, could highlight the value of human skill and creativity. Cultural initiatives like this reinforce that while AI is becoming ever-present, society is still interested in living human artists and their irreplaceable societal role as creators. Economically, the result is a "handmade movement" for art, where the artworks have been verified as human-made, perhaps with blockchain certificates or something similar to differentiate it from any AI-generated version of it, could be perceived as having a specific premium value. The recommendation is for cultural institutions to be prepared for this and possibly proactively promote human-only art spaces or grants as a measure to keep the artistic ecosystem in balance.

### 6.6 *Continuous research on AI's impact and adaptation*

Ultimately, we should continue to research the impact of AI on the arts as AI continues to develop and expand. Academia and research organizations can provide us with longitudinal research to understand how artists' careers are being impacted by AI, how different art forms are affected by AI (is some use of AI more transformative than others?), and what opportunities for new careers in the arts have arisen. This research can help to inform either policy or education initiatives. Also, keep an eye on any unintended consequences (for instance, if certain art domains are suffering more than benefiting from AI) that may require mid-course correction. AI is rapidly developing, making it important to assess anything iteratively. Most importantly, if there are any positive case studies - success stories where an artist used AI to create something novel or that an artist used AI to solve a problem about preservation - then those should be circulated as much as possible, as blueprints for the broader adoption of AI in the arts.

Combining these future directions, we can begin to recognize an art world where AI is used responsibly. Ultimately, the balancing act is about the creative possibilities that AI offers to be more efficient in a variety of ways, and the human imperatives embedded in the arts and art-making. That future would be one of "inclusive innovation", where AI could serve as a bridge, not a barrier, between the future of the arts, the legacy of tradition, and technology in the arts [8].

## 7. Conclusion

The impact of AI on traditional art forms is certainly significant, but it is a complicated impact and certainly not a pure positive or negative. As we have seen in this paper, AI can be both a negative disruption and a positive enhancement, and often both at the same time. The conversations long-time, established art forms have with artists and society must make us rethink creativity, authorship, and value. AI can introduce uncertainty about jobs and authenticity [4, 1]. On the other hand, AI creates amazing possibilities - painters creating with the help of algorithms, choreographers making movements with the help of suggestions from a machine, and craftsman reinventing outputs with generative models [3, 8]. The two examples and evidence we have looked at illustrate that the contribution of AI is situational and specific. If artists do not apply AI in the correct ways and limit themselves to AI generative models, some traditional art forms may lose part of their cultural and economic basis, a true disruption in a negative manner. If the artist employs AI with a thoughtful approach, then AI is simply extending its reach and preserving what was not previously sustainable, and that is a true companion to human creativity. As we look forward to figuring out what the future of AI means for art, the future of AI will be directed by more than just the technology, but also by our capacity to make decisions as humans about how we use and embed that technology as a society. The onus is on the art community, technologists, and policymakers to determine how, together, we can create ways of navigating the future that develop, further, or sustain an ethical standard that allows human creativity to flourish in conjunction with the increasing scale of AI in the future. This includes equipping artists with relevant resources and training to learn new ways of making, establishing a fair practice on generative models, access to artistic material, and educating consumers on the shifting nature of what we understand art to be. And while we begin to undertake these changes, it is worth noting that historically, art forms have adopted or at the very least engaged with new technologies, from the camera to digital-based software, never to remove the artists but to provide additional ways of expressing. While AI could be seen as one of the most impactful technologies so far, AI is simply one tool. The emotion, the experience, and the cultural nature of art still reside within the human experience. If we can continue to place this truth in the centre, even if AI disturbs some aspects of the art with a zero-sum game, even so, AI could serve as a catalyst, together with human creativity promise to elevate art above what is seen today.

## 8. Conflict of Interest

The authors have no conflicts of interest to declare.

**Authors Profile**

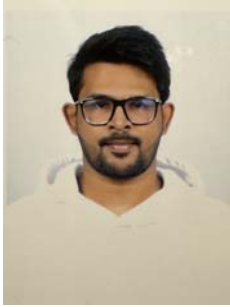

**Viswa Chaitanya Marella**, is a Data Analyst with a Master's degree in Data Analytics. He combines technical precision with a passion for solving complex challenges through data-driven insights. His interests span research analysis, automation, machine learning algorithms, and applications of AI. As an AI enthusiast, Viswa is dedicated to continuous learning and advancing his skills in data analysis and AI.

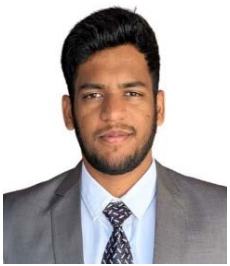

**Sai Teja Erukude**, is a seasoned Data Scientist and passionate researcher with over five years of experience in the field. Sai holds a Master's degree in Computer Science with a specialization in Data Science. He brings a solid blend of academic knowledge and hands-on expertise, with a focus on pushing the boundaries of AI. His areas of interest include Generative AI, Deep Learning, and real-world AI applications. Guided by a strong sense of purpose, Sai is dedicated to make positive impact through meaningful research.

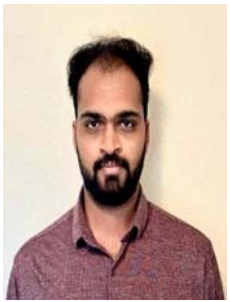

**Suhasnadh Reddy Veluru**, is a Data Engineer with a Master's degree in Data Analytics and Engineering. Suhas blends technical precision with a passion for solving complex problems using data-driven approaches. His interests span across data architecture, automation, and the application of AI in real-world scenarios. As an AI enthusiast, Suhas is committed to contributing to impactful, forward-thinking projects that bridge the gap between data infrastructure and intelligent systems.